\documentstyle[eqsecnum,aps]{revtex}
\begin{document}
\newlength{\letw}
\draft

\title{Relativistic description of
electron scattering on the deuteron}
\author{E. Hummel$\,{}^*$ and J.\ A.\ Tjon}

\address{ Institute for Theoretical Physics, University
of Utrecht,\\
 Princetonplein 5, 3508 TA Utrecht, The Netherlands}

\maketitle

\begin{abstract}
Within a quasipotential framework a relativistic analysis is
presented of the deuteron current. Assuming that the
singularities from the nucleon propagators are important, a so-called
equal time approximation of the current is constructed. This is
applied to both elastic and inelastic electron scattering.
As dynamical model the relativistic one boson exchange model is
used.
Reasonable agreement is found with a previous relativistic
calculation of the elastic  electromagnetic form factors of the
deuteron.  For the unpolarized inelastic electron scattering
effects of final state
interactions and relativistic corrections to the structure functions
are considered in the impulse approximation.
Two specific kinematic situations are studied as examples.
\end{abstract}
\vspace{2cm}

\pacs{PACS numbers: 11.10.Qr, 25.30-c, 25.30.Bf}

\section{INTRODUCTION}

Electron scattering processes have been crucial in extracting
detailed information about the nuclear interaction. Various
exclusive \cite{mvds,vdst,ducret} and inclusive \cite{Parker,Dytman}
breakup reactions for the few nucleon systems have recently for
example been performed at moderate momentum transfer,
where separations of the longitudinal and transverse
response functions have successfully been achieved.
The deuteron, as the most simple nuclear system, is of special
interest because  exact calculations are in principle feasible
for this case. Such a system may  serve as testing ground for
theoretical models describing the nuclear dynamics and the e.m.
operators used to study these reaction processes.
In most of these studies it is implicitly assumed that the
constituents of the system behave nonrelativistically and that
the e.m. interaction can be treated in an independent way. With
increasing momentum  and energy transfer effects of relativity
are expected to play an important role and a relativistic
description should be essential in these kinematic regions. Moreover,
in view of gauge invariance a consistent treatment
of the e.m. interaction is called for.

In this paper we discuss a framework which
is well suited for a relativistic analysis of
both the elastic and inelastic electron scattering off
the deuteron. It is based on the relativistically
covariant field theoretical Bethe-Salpeter equation approach.
Gauge invariance is satisfied through the Ward-Takahashi
identity. Assuming a theoretical description in terms of
nucleonic and mesonic degrees of freedom the elastic
e.m. properties have already been studied. Within an one boson
exchange (OBE) model \cite{FlTj} a consistent relativistic treatment of
both the e.m. current and the
nucleon-nucleon interaction can be realized \cite{ZuTj80},
including also meson exchange current( MEC) contributions from
the $\rho \pi \gamma$ and $\omega \varepsilon \gamma$ currents
\cite{hutj}.
In the actual calculations a relativistic quasipotential
approximation of the Blankenbecler-Sugar-Logunov-Tavkhelidze
(BSLT) has been used.

The disintegration experiments of the deuteron yields
additional information as compared to the elastic
e.m. scattering process. A larger kinematic region can be
tested and the theoretical models are more complicated due to final
state interaction (FSI) and other MEC contributions as is the case in
elastic scattering. Another point of interest is that, depending
on the type of experiment, the cross section can be separated in
more than 2 structure functions. The inclusive experiment is
described by the longitudinal $R_L$ and transverse $R_T$ structure
functions. For the exclusive experiment studied here an
additional structure function $R_{TL}$ is measured, which is an
interference between the longitudinal and transverse components in
the current operator. Because this structure function depends on
other reaction processes as $R_L$ and $R_T$, new information can
be obtained.

In our calculations complete knowledge of the FSI and the
deuteron wave function is required. As mentioned earlier we use
the OBE model of Ref. \cite{FlTj}.
Although the complete model can be studied we use here
a quasipotential approximation to it, where both particles are
treated in a symmetrical way. This BSLT approximation
\cite{BSLT,lt} was also used in the model of the elastic form
factors. Here
we'll use the same approximation for the initial and final
state, but use a different approximation for the nucleon
propagators occurring in the e.m. vertex.
We assume that only a relative energy dependence is in these
propagators and not in the two-nucleon vertex functions.
As a result the relative energy dependence  can explicitly
be integrated out. Because the BSLT  choice is used for the
nucleon states this is an equal time (ET) approximation
to only the current operator.
To examine the sensitivity of the predictions on such a
ET prescription for the e.m. current matrix elements we
calculate the e.m. form factors in the elastic case.
Comparing to earlier calculations \cite{hutj} we find
only small differences. The ET
choice has the advantage that it can also be extended in
a systematic way to
describe the case of inelastic electron scattering.  In the full
Bethe-Salpeter theory a conserved deuteron current can be
constructed (see Refs. \cite{ZuTj80,GrRi}),
at least for the case of strong form factors at the
meson-nucleon vertices, which depend only on the momentum of the
meson.  The same can be achieved in a systematic way in the ET
approximation, i.e. MEC contributions can be
constructed for the breakup reaction to ensure current
conservation.

The paper is organized as follows. In the next two sections
we summarize the relevant expressions, characterizing
the inelastic electron scattering processes and
describe the dynamical model used in this work.
In particular, we present a new fit in this model to the
experimental phase shifts, to discuss the effect of
including the negative energy states. In sections IV
and V the relativistic impulse approximation and the
elastic e.m. form factor calculations are described in
the ET approximation. Section VI deals with the
formalism of the e.m. breakup of the deuteron in this
approximation. In all these calculations a fully
relativistic e.m. current operator with on shell
form factors is used. In section VII two kinematic
situations are studied using this relativistic
formalism.
For the conventional nonrelativistic (NR) models different
forms of e.m. operators have been constructed in the literature.
These models differ predominantly in the use of $F_1$ or $G_E$
for the e.m. nucleon form factors and
the differences are obviously of relativistic order.
We compare these predictions with our fully relativistic analysis and
show that in the kinematics studied the predictions
can be very different. Most of the differences can be
traced back to the choice of the e.m. operator.
Although the experiments are at moderate momentum
transfer, large relativistic effects are in particular
found in the interference structure function $R_{TL}$,
showing the failure of the NR models discussed here.
The next leading order correction to these operators is given
and shown to represent reasonably well the relativistic predictions,
at least in the kinematics considered.
In the kinematic situations discussed in this section
the MEC contributions which
are needed to ensure gauge invariance are of minor
importance. In a next paper we will discuss in detail
our MEC calculations in this relativistic model.

\section{DEUTERON STRUCTURE FUNCTIONS}

In this section we summarize some useful formulae describing the
electrodisintegration process of the deuteron.
The conventions of Bj\"{o}rken and Drell \cite{BjDr} are used.
We confine ourselves in this paper to the description of unpolarized
electron scattering on an unpolarized target and follow closely
the work of Donnelly and Raskin\cite{DoRa}. The analysis can readily be
extended to the case of polarized scattering.

For definiteness, let us consider the breakup of a
deuteron with total momentum $P$ by an electron into a free
neutron-proton (np) pair, characterized by the four momenta
$p_n = ({\bf p}_n, E_n)$ and $p_p = ({\bf p}_p, E_p)$ respectively.
In the one-photon exchange approximation, the
differential cross section in the lab system, being the rest
frame of the deuteron i.e. $P =({\bf 0}, M_D)$,
can be written as \cite{BjDr}
\begin{equation}\label{cross}
d \sigma = C_{np} \frac{\varepsilon}{|{\bf k}|} \sum_{if}^{-}
\frac{m_e^2}{\varepsilon'\varepsilon}\frac{d^3 k'}
{(2 \pi)^5}
d^3 p_p d^3 p_n
\delta^4(k+P-k'-P')\frac{e^4}{q_\mu^4} |j_\mu^e J^\mu|^2,
\end{equation}
with $C_{np}= M_N^2/(E_n E_p)$ and where $j_\mu^e$ is the
electron
and $J^\mu \equiv <f \mid J_d^{\mu} \mid i >$
the deuteron electromagnetic current matrix element
between the initial and final state of the nuclear system.
The total four momentum of the np pair is assumed
to be given by $P'$, while
$q$ is the four momentum
of the virtual photon satisfying $q_\mu  = k_\mu
-k'_\mu \equiv ({\bf q},\omega)$,
$k_\mu = ({\bf k},\varepsilon)$  and $k_\mu' = ({\bf k}',\varepsilon')$
being the four momenta of the incoming and scattered electron.
Momentum conservation at the
photon-deuteron vertex gives $P+q=P'$.
The normalization of the outgoing np pair state is such that
when the final state interaction is neglected it is given by
$<p_p,p_n \mid f> \approx
u({\bf p}_p) u({\bf  p}_n)$,
$u$ being Bj\"{o}rken-Drell spinors.
In Eq.~(\ref{cross})
${\displaystyle \sum_{if}^{-}}$
indicates  averaging and summing over
both the electron and nuclear polarizations in the initial and final
state respectively.
An electron and hadronic current tensor
\begin{equation}\label{etens}
\eta_{\mu \nu}  =
\sum_{if}^{-}
j_\mu^{e^*} j_\nu^e,\;
W_{\mu \nu}  =   C_{np}\sum_{if}^{-} J_\mu^* J_\nu,
\end{equation}
can now be introduced such that
${\displaystyle \sum_{if}^{-}|j_\mu^e J^\mu|^2 = \eta_{\mu \nu} W^{\mu \nu}
}$.
The expression for $\eta_{\mu \nu}$ is well known\cite{BjDr} and
is given in the ultra relativistic limit ( $m_e \approx 0$) by
\begin{eqnarray}\label{etens3}
\eta_{\mu \nu}
 & = & \frac{1}{2 m_e^2} ( k_\mu k_\nu' + k_\nu k_\mu' -g_{\mu
\nu}  k\cdot k'  ).
\end{eqnarray}
Introducing the relative
momentum ${\bf p}'= ({\bf p}_p- {\bf p}_n)/2$ of the np pair
and integrating over ${\bf P}'$ yields for the cross section
\begin{equation}\label{cross2}
d \sigma = \frac{m_e^2}{|{\bf k}| \varepsilon'}
\frac{e^4}{q_\mu^4} \frac{d^3 k' d^3 p'}
{(2 \pi)^5}
\delta(M_D+\omega-E_p-E_n) \eta_{\mu \nu} W^{\mu \nu}.
\end{equation}
At this point the concept of the structure functions can be
introduced.  We may choose a coordinate system, where ${\bf q}$
is along the $z$-axis, while the
$y$-axis is defined along ${\bf k} \times {\bf k}'$.
Let us now define in this lab frame
$J_\pm = \mp \frac{1}{\sqrt{2}} \left( J_x \pm i J_y \right)$,
and use current conservation $q \cdot J = 0$ to eliminate the
longitudinal third component $J_3$ of the current in favor of the
charge $J_3 = \omega/q J_0$.
With the help of Eq.~(\ref{etens3}) the contraction can be written in
the form
\begin{equation}\label{contr}
\eta_{\mu \nu} W^{\mu \nu} = v_0 \left(  v_L R_L + v_T R_T + v_{TT}
R_{TT} + v_{TL} R_{TL} \right),
\end{equation}
with $v_0 = 4 \varepsilon\varepsilon' \cos^2 \frac{1}{2} \theta_e$
, $\theta_e$ being the scattered electron angle.
Moreover, the electron-kinematic factors are given by
$v_L  =  \left( \frac{q_\mu^2}{{\bf q}^2} \right)^2$,
$v_T=-\frac{1}{2}\frac{q_\mu^2}{{\bf q}^2}+\tan^2 \frac{1}{2}\theta_e$,
$v_{TT} = \frac{1}{2}\frac{q_\mu^2}{{\bf q}^2}$
and
$v_{TL} = \frac{1}{\sqrt{2}}\frac{q_\mu^2}{{\bf q}^2}\sqrt{-
    \frac{q_\mu^2}{{\bf q}^2}+ \tan^2 \frac{1}{2}\theta_e}$.
The nuclear structure functions $R_\alpha$ can be expressed in the
matrix elements of the various current components as\cite{DoRa}
\begin{eqnarray}\label{Rk}
R_L   &=&   C_{np} |J_0|^2 ,
\nonumber \\
R_T   &=&   C_{np} ( |J_+|^2 + |J_-|^2 ),
\nonumber \\
R_{TT}  &=&   2 C_{np} Re \left(J^*_+ J_-\right) ,
\nonumber \\
R_{TL} &=& -2 C_{np} Re \left[ J^*_0 \left( J_+ - J_- \right) \right].
\end{eqnarray}
It should be noted that $C_{np}$ can be absorbed in the
Bj\"{o}rken-Drell spinors
occurring in the e.m. current matrix elements. The resulting spinors
become essentially those defined by Kubis \cite{Ku}, described in
Appendix \ref{helicity}.
In the remainder of the paper these spinors will be
used.
Introducing the Mott cross section
$\sigma_{Mott} = \frac{\alpha^2}{q_\mu^4} \frac{\varepsilon'}{
\varepsilon}v_0$,
with the fine structure constant given by $\alpha = e^2/(4
\pi)$, the cross section can be written as
\begin{equation}\label{cross3}
d \sigma =   \sigma_{Mott}
\left(  v_L R_L + v_T R_T + v_{TT} R_{TT} + v_{TL} R_{TL} \right)
\delta(M_D+\omega-E_p-E_n) d\varepsilon'd\Omega_e' d^3p'.
\end{equation}
This is still a general expression, except that current conservation
of the hadronic current is assumed.

All the quantities in Eq.~(\ref{cross3}) are determined in the
lab frame.  Considering the total four momentum of the np pair
$P' = ({\bf q}, E_{np})$ (with
${\bf q} = {\bf p}_n + {\bf p}_p$ and $E_{np} = E_n + E_p$)
an invariant mass $M_{np}$ of the pair can be defined as $E_{np} =
\sqrt{{\bf q}^2 + M_{np}^2}$.
Since the final state interaction of the final np pair can most
readily be determined in its c.m. system
($P'^{cm} = ( {\bf 0}, M_{np})$),
we prefer to express Eq.~(\ref{cross3})
in terms of the structure functions $R_\alpha^{cm}$, evaluated
in this frame. To make a distinction between lab and c.m.
variables we use the convention that all variables in the
c.m. frame are labeled by $cm$.
The relation can easily be found from
\begin{equation}\label{lor}
J_\mu^{cm} = {\cal L}^{\mbox{ } \nu}_\mu J_\nu.
\end{equation}
where ${\cal L}^{\mbox{ } \nu}_\mu$
describes the boost between the lab and c.m. frame.
An explicit expression for the
Lorentz transformation can be found using the relation
$P'^{cm}_\mu = {\cal L}^{\mbox{ } \nu}_\mu P'_\nu$ with
\begin{equation}\label{lortr}
 {\cal L}_{\mu}^{\mbox{ } \nu} =
 \left( \begin{array}{cccc}
   \sqrt{1+\eta} & 0 & 0 &  - \sqrt{\eta} \\
   0 & 1 & 0 & 0 \\
   0 & 0 & 1 & 0 \\
    - \sqrt{\eta} & 0 & 0 & \sqrt{1+\eta}
  \end{array}  \right),
\end{equation}
where we have
\begin{equation}\label{eta}
\sqrt{\eta}  =  \frac{q}{M_{np}},\;
\sqrt{1+\eta}  =  \frac{E_{np}}{M_{np}}.
\end{equation}
Since ${\bf q}$ is along the z-axis, the transverse components of
the deuteron current are not affected by the Lorentz boost. For the
charge component we get
$J^{cm}_{0} = \sqrt{1+\eta} J_0 - \sqrt{\eta} J_3$.
Together with current conservation
$q \cdot J = q^{cm} \cdot J^{cm} = 0$,
we find a direct relation between the charge components of the
current in the two frames
\begin{equation}\label{j0cm2}
J_0 = \frac{M_{np}}{M_{D}} J_0^{cm}.
\end{equation}
We now turn to derive the differential cross section describing
the exclusive $d(e,e'p)n$ and inclusive $(e,e')$ reactions.
In view of the Bj\"orken and Drell covariant normalization of
the free np spinors,  the transformation property of the
$C_{np}$ coefficient and the invariant volume element
$ d^3p'^{cm}/E_{p'}^{cm} = d^3p'/E_{p'}$ we may rewrite
Eq.~(\ref{cross3}) as
\begin{equation}\label{cross4}
d \sigma  =  \sigma_{Mott}
\left(  v_L R_L + v_T R_T + v_{TT} R_{TT} +
   v_{TL} R_{TL}\right)
\delta(E_D^{cm}+\omega^{cm}- 2 E_{p'}^{cm}) d\varepsilon'd\Omega_e'
d^3p'^{cm},
\end{equation}
where the structure functions can be expressed in the
corresponding observables in the c.m. system
\begin{equation}\label{rlab}
R_L =   \left( \frac{M_{np}}{M_{D}} \right)^2 R_L^{cm} ,\;
R_T =   R_T^{cm},\;
R_{TT} =   R_{TT}^{cm} ,\;
R_{TL} =   \left( \frac{M_{np}}{M_{D}} \right) R_{TL}^{cm}.
\end{equation}
The exclusive reaction $d(e,e'p)n$ is represented by  a five
fold differential
cross section.
In view of the invariant volume element
we immediately find from
Eq.~(\ref{cross4})
\begin{equation}\label{exclusive}
\frac{d^5 \sigma }{d \varepsilon' d\Omega_e' d\Omega'} =
\sigma_{Mott}
\left(  v_L {\hat R}_L + v_T {\hat R}_T
+ v_{TT} {\hat R}_{TT} + v_{TL} {\hat R}_{TL} \right),
\end{equation}
where we have absorbed the Jacobian $J$ into the definitions of the
exclusive structure functions ${\hat R}_\alpha \equiv
J R_\alpha$.
For the Jacobian we have
$J = \frac{d \Omega'^{cm}}{d \Omega'} \frac{1}{2} p'^{cm} E^{cm}_{p'}$
with
\begin{equation}\label{jacobian}
 \frac{d \Omega'^{cm}}{d \Omega'}  =
\frac{p'}{p'^{cm}}  \frac{d E_{p'}}{d E^{cm}_{p'}}
 =  \frac{p'}{p'^{cm}} \frac{1}{\sqrt{1+\eta}}
\left( 1 -
\sqrt{\frac{\eta}{1+\eta}} \frac{E_{p'}}{p'} \cos
\theta' \right)^{-1}_.
\end{equation}
Hence, in principle the unpolarized exclusive reaction can be
calculated by determining the four structure functions.
For the $(e,e')$ process the integration over the momenta of
the final np pair has to be performed. This is actually done in
the c.m. frame of the  np pair.
Integration over the relative momentum ${\bf p}'^{cm}$
yields the inclusive $(e,e')$ cross section (in the lab frame)
\begin{equation}\label{inclusive}
\frac{d^3 \sigma}{d \varepsilon' d\Omega_e'} =
 \sigma_{Mott} \left( v_L \overline{R}_L+ v_T
\overline{R}_T\right),
\end{equation}
where the inclusive structure functions $\overline{R}_\alpha$ are
defined by
\begin{equation}\label{inclusiv1}
\overline{R}_\alpha = 2 \pi \int_{-1}^{1}  R_\alpha
\frac{1}{2} p'^{cm} E_{p'}^{cm} d \cos \theta'^{cm}.
\end{equation}
The momentum $p'^{cm}$ is determined by the energy conserving $\delta$
function in Eq.~(\ref{cross4}), i.e. from
$E_{p'}^{cm} = \frac{1}{2} \left( E_D^{cm} + \omega^{cm} \right)$.
The contributions to the inclusive cross section from $R_{TT}$ and
$R_{TL}$, which are proportional to $\cos 2 \phi'^{cm}$ and  $\cos
\phi'^{cm}$ respectively, vanish due to the $\phi'^{cm}$-integration.
In section VI we
will discuss this $\phi'^{cm}$ dependence in more detail.

\section{THE FORCE MODEL}
In determining the nuclear structure functions we in principle
need to know the deuteron vertex function and the
half off-shell nucleon-nucleon (NN) t-matrix.
The nuclear interaction used by us is
based on the one-boson-exchange (OBE) model.
Following Fleischer and Tjon \cite{FlTj} it is assumed to be
described by the exchange of
$\pi, \rho, \omega, \eta, \varepsilon$ (or $\sigma$) and
$\delta$ mesons.
A strong meson-nucleon form factor of the monopole type
$F(p^2) = \frac{\Lambda^2}{\Lambda^2-p^2}$
is used to regularize the large momentum behavior.
Within the relativistic field theory two-particle scattering can
be described by the scattering t-matrix $\phi$, which satisfies
the Bethe-Salpeter equation.
The inhomogeneous Bethe-Salpeter equation for the t-matrix
has the form
\begin{eqnarray}\label{bsin}
 \phi(p',p;P)  &=&  V(  p',p)
 -\frac{i}{4 \pi^3} \int d^4k \phi(p',k;P)
 S_2(k,P) V(k,p)
 \nonumber \\
 &=&  V(  p',p)
 -\frac{i}{4 \pi^3} \int d^4k V(p',k) S_2(k,P) \phi(k,p;P),
\end{eqnarray}
where
\begin{eqnarray}\label{s2}
S_2(p,P) & = & S^{(1)}(p,P) S^{(2)}(p,P)
\nonumber \\
 &=&  (\frac{1}{2}
{P \settowidth{\letw}{$P$}
 \hspace{-0.4\letw}
 \makebox[0cm]{/}
 \hspace{0.4\letw}}
+
{p \settowidth{\letw}{$p$}
 \hspace{-0.4\letw}
 \makebox[0cm]{/}
 \hspace{0.4\letw}}
+M_N)^{(1)}
 (\frac{1}{2}
{P \settowidth{\letw}{$P$}
 \hspace{-0.4\letw}
 \makebox[0cm]{/}
 \hspace{0.4\letw}}
 -
{p \settowidth{\letw}{$p$}
 \hspace{-0.4\letw}
 \makebox[0cm]{/}
 \hspace{0.4\letw}}
+M_N)^{(2)} G_0,
\end{eqnarray}
with
$G_0 = [(\frac{1}{2} P + p)^2 - M_N^2 + i \varepsilon]^{-1}
[(\frac{1}{2} P - p)^2 - M_N^2 + i \varepsilon]^{-1}$.
In Eq.~(\ref{s2}) $S^{(n)}(p,P)$ is the free propagator of the
n-th nucleon.
The t-matrix also determines
the deuteron vertex function $\Phi_D^{(M)}$ (p,P) (where M is the
polarization of the deuteron), corresponding to the residue at the
deuteron bound state pole at $P^2 = M_D^2$. For $P^2 \approx
M_D^2$ we have
\begin{equation}\label{pole}
\phi(p',p;P) =  \sum_M \frac{\Phi_D^{(M)}(p',P)
\tilde{\Phi}_D^{(M)} (p,P)}{P^2-M_D^2}
+ \mbox{regular terms}.
\end{equation}
Alternatively, $\Phi_D$ satisfies the homogeneous equation
\begin{equation}\label{bshom}
\Phi_D^{(M)} (p,P) = - \frac{i}{4 \pi^3}\int d^4 k V(p,k) S_2(k,P)
\Phi_D^{(M)} (k,P).
\end{equation}

Although a field theoretical Bethe-Salpeter analysis is in principle
possible for the case that $V$ is given by the OBE model,
calculations are highly non trivial because of its
analytic structure and computational complexity.
To simplify this we  use a  relativistic quasipotential approximation.
In the quasipotential framework the two-particle propagator is
replaced by one where the relative energy variable is constrained
(for a review see \cite{Tjlh}). Here
we use the choice which treats the two nucleons in a symmetric way
\cite{BSLT,lt}. It is given by
\begin{equation}\label{tnqp2}
G_0 \rightarrow i \pi \delta(p_0) G^{BSLT}_2 = i \pi
\delta (p_0) \frac{1}{E_p -E} \frac{1}{(E_p+E)^2},
\end{equation}
where $E = \frac{1}{2} P_0$ and $E_p = \sqrt{M_N^2 + {\bf p}^2}$.
In this approximation the full
two-particle propagator $S^{BSLT}_2$, containing the spinor structure
(see Appendix \ref{helicity}),
is given by
\begin{equation}\label{sqp}
S_2^{BSLT}  = (\frac{1}{2}
{P \settowidth{\letw}{$P$}
 \hspace{-0.4\letw}
 \makebox[0cm]{/}
 \hspace{0.4\letw}}
+
{p \settowidth{\letw}{$p$}
 \hspace{-0.4\letw}
 \makebox[0cm]{/}
 \hspace{0.4\letw}}
+M_N)^{(1)}
 (\frac{1}{2}
{P \settowidth{\letw}{$P$}
 \hspace{-0.4\letw}
 \makebox[0cm]{/}
 \hspace{0.4\letw}}
 -
{p \settowidth{\letw}{$p$}
 \hspace{-0.4\letw}
 \makebox[0cm]{/}
 \hspace{0.4\letw}}
 +M_N)^{(2)} G^{BSLT}_2
 =  (E_p - E) S^{(1)} (p,P)  S^{(2)} (p,P).
\end{equation}
\noindent
In this approximation the inhomogeneous Bethe-Salpeter equation
(\ref{bsin})
simplifies to
\begin{equation}\label{bsinqp}
 \phi({\hat p},{\hat p}';P)  =  V({\hat p},{\hat p}')
	 +\frac{1}{4 \pi^2} \int d^3k
           \phi({\hat p},{\hat k};P)
             S_2^{BSLT}({\hat k},P) V({\hat k},{\hat p}'),
\end{equation}
where ${\hat p}, {\hat p}'$ and ${\hat k}$ are the four vectors $p, p'$
and $k$, subject to the condition that their fourth component
vanishes in the two-nucleon c.m. frame.
To reconstruct the off-shell scattering wave function needed in
the study of FSI effects a partial wave representation is used
for the full NN amplitude. Following Ref.~\cite{FlTj}, the helicity
basis is used. The representation is briefly
discussed in Appendix \ref{helicity}.
The on-shell amplitude is simply obtained by taking $p=p'$ with
$\sqrt{P^2} \equiv \sqrt s =2 E_p$.
Eq.~(\ref{bsinqp}) can be solved in a partial wave analysis.
For details we refer to \cite{FlTj}.
Because the quasipotential approximation is used, Eq.~(\ref{bsinqp})
reduces essentially to a coupled set of one-dimensional integral
equations. Besides the physical $(+,+)$ positive energy states,
also $(-,-)$ states and  even and odd combinations of
the $(+,-)$ energy states occur \cite{Ku} .

Starting from the original fit of Ref. \cite{FlTj} (set A) for only
positive energy spinor states, the meson-nucleon coupling constants
$g_\epsilon$ and $g_\delta$ were adapted to the case that all the spinor
states are included to get a reasonable fit to the experimental phase
shifts of Arndt et al \cite{Arndt}.
The resulting phase shifts up to $J = 2$ are shown in Fig.~\ref{nnph}
(set B), where the spectroscopic notation ${}^{2S+1}L_J$ has been used.
For comparison the results of fit A are also plotted. To see
what the effects are of the negative energy spinor states, we
have switched these states off, keeping the same strength for
the coupling constants. The results are shown as
the dotted-dashed lines.
Finally, an attempt was made to vary the meson coupling
constants for the case of only positive energy states to
reproduce the phase shifts as obtained with all spinor states
included. We indeed obtain phase shifts (set C) which are not
distinguishable from the set B in fig.~\ref{nnph}.
The sets of coupling parameters are given in Table I.

\section{IMPULSE APPROXIMATION}

In order to construct a quasipotential approximation
for inelastic electron scattering we start with
the current operator as given in the Bethe-Salpeter formalism.
The contributions to the current operator in the impulse
approximation (IA) are given
by the Feynman graphs shown in Fig.~\ref{pwiad} and
Fig.~\ref{fsid} in the case of
inelastic electron  scattering. In the
contributions (a) and (b) ((c) and (d)) in Fig.~\ref{pwiad} the photon
couples directly to the proton (neutron) which is knocked out without
any interaction with the other nucleon. These contribution are called
the plane wave impulse approximation (PWIA) and the Born term
respectively. The set of graphs in Fig.~\ref{fsid} corresponds
to the contributions, where the outgoing nucleons
interact after the virtual photon has been absorbed. These are the
so-called final state interaction (FSI) contributions. In our
calculations we will absorp the Born contribution in the FSI
contributions.
As is seen from the figure the FSI is expressed
in terms of the half-off-shell
NN t-matrix $\phi$. It  satisfies the inhomogeneous equation
given in Eq.~(\ref{bsin}). Including the Fermi character of the
nucleons, the free np pair with total
momentum $P'$ and relative momentum $p_f$ is given by the
antisymmetric combination of two free Dirac spinors  with
helicities $\lambda_n$
\begin{equation}\label{nppair}
 < p_f,P'| =
 \overline{u}_{\lambda_{1}} ( \frac{1}{2} {\bf P}' +{\bf p}_f)
\overline{u}_{\lambda_{2}} ( \frac{1}{2} {\bf P}' -{\bf p}_f)  A,
\end{equation}
where A is the antisymmetrizer.
As mentioned before we use the Kubis spinors.
Since the outgoing particles are on-shell, we have to satisfy
the two conditions
\begin{equation}
p_f \cdot P' = 0,\;
P'_0 = E_{\frac{1}{2} {\bf P}' +
{\bf p}_f} + E_{\frac{1}{2} {\bf P}' - {\bf p}_f}.
\end{equation}
The PWIA and FSI contributions can be combined by introducing
the scattering wave function for the two nucleons
\begin{equation}\label{scat}
{\tilde \Psi}_{np} (p_f,p';P') =  <p_f,P'| \; [ 4 \pi^3 i
\delta^4 (p'-p_f)
+  {\tilde \phi_{np}}(p_f,p';P') S_2(p',P') ],
\end{equation}
where ${\tilde \phi_{np}}$ satisfies Eq.~(\ref{bsin}), subject to
the condition that both outgoing particles are on mass shell.
As a result, the current operator in the IA as
represented in Figs.~\ref{pwiad} and \ref{fsid} can be written in the form
\begin{equation}\label{bsia2}
J_\mu^{IA}  =  \frac{ i e}{(2 \pi)^4 M_D} \sum_{i=1}^2
\int d^4p'{\tilde \Psi}_{np}(p_f,p';P')
\Gamma_\mu^{(i)}(q)S^{(i)}(p^{(i)},P)\Phi_D (p^{(i)},P),
\end{equation}
where
$p^{(1)} = p' -\frac{1}{2} q$ and $p^{(2)} = p' + \frac{1}{2} q$.
Because of four momentum conservation we have $P'=P+q$.
For the $\gamma NN$
vertex for the k-th nucleon we take
\begin{equation}\label{gnn}
\Gamma_\mu^{(k)}(q) = F_1^{(k)}(q) \gamma_\mu^{(k)} +\frac{i}{2 M_N}
F^{(k)}_2(q) \sigma_{\mu \nu}^{(k)} q^\nu,
\end{equation}
with $F^{(k)}_n= F_n^S+\tau_3^{(k)} F_n^V$.
In the actual studies we assume that the e.m. nucleon form factors
$F_n$ can be described by their on-shell form.
To obtain a similar form for the current matrix element as in the
elastic case we may introduce the corresponding scattering
vertex function
\begin{equation}\label{vert}
{\tilde \Phi}_{np} (p,p';P)={\tilde \Psi}_{np} (p,p';P) S_2^{-1}(p',P).
\end{equation}
{}From studies of the two-nucleon functions $\Phi_{np}$ and
$\Phi_D$ \cite{thesis} a rather smooth behavior is found in the
relative energy $p_0$ variable. Therefore we assume that
an expansion of the vertex functions
in Eq.~(\ref{bsia2}) around their relative variable $p_0=0$
point is reasonable in practical calculations. Keeping only
the lowest order contribution and assuming that the
resulting $\Phi's$ can be taken to be the BSLT vertex function,
we get for the e.m. current operator
\begin{equation}\label{loopqp2}
 J_{\mu}^{IA}  = \frac{ie}{(2 \pi)^{4} M_D} \sum_{i=1}^2 \int d^{4}p'
 {\tilde \Phi}_{np}^{BSLT}({\hat p}';P') S_{2}(p',P')
\Gamma_{\mu}^{(i)} S^{(i)}(p^{(i)},P) \Phi_D^{BSLT}({\hat p}^{(i)};P).
\end{equation}
Since the only $p_0$ dependence is in the nucleon propagators
we may carry out the $p_0$ integration analytically.
In Appendix  \ref{k0int} the explicit formulae of the propagator
structure are given. The above
approximation is essentially an equal time (ET) approximation to the
current operator, i.e. with zero relative time.

Let us consider the covariance aspects of the quasipotential
description.
Under a given Lorentz transformation ${\cal L}$ we have
\begin{eqnarray}\label{lor2}
V(k,p)  &=&  \Lambda({\cal L}) V( {\cal L}^{-1}k, {\cal L}^{-1}p)
\Lambda^{-1} ({\cal L}) ,
\nonumber \\
S^{(i)}(p,P)  &=&  \Lambda^{(i)}({\cal L}) S^{(i)}( {\cal L}^{-1}p,
{\cal L}^{-1}P) \Lambda^{{(i)}^{-1}}({\cal L}),
\nonumber \\
\Phi_D(p,P)  &=&  \Lambda({\cal L}) \Phi_D( {\cal L}^{-1}p,
{\cal L}^{-1}P) ,
\nonumber \\
\phi_{np}(k,p,P)  &=&  \Lambda({\cal L}) \phi_{np}( {\cal L}^{-1}k,
{\cal L}^{-1}p, {\cal L}^{-1}P) \Lambda^{-1} ({\cal L}),
\nonumber \\
\end{eqnarray}
with
$\Lambda({\cal L}) = \Lambda^{(1)}( {\cal L})\Lambda^{(2)}( {\cal L})$,
where  $\Lambda^{(i)}({\cal L})$ is the corresponding boost
operator in the Dirac space of the i-th nucleon.
Because the BSLT deuteron and continuum wave functions transform in a
similar way as the Bethe-Salpeter ones, the transformation property
of the deuteron current for the quasipotential case is the same
as that of the relativistic field theory case. Since
\begin{equation}
\Lambda^{{(1)}^{-1}}({\cal L})\Gamma_\mu^{(1)}(q) \Lambda^{(1)}( {\cal
L}) = {\cal L}_\mu^{\mbox{ } \nu} \Gamma_\nu^{(1)}({\cal L}^{-1} q),
\end{equation}
we may conclude that the deuteron current given by
Eq.~(\ref{loopqp2}) indeed transform as a four vector.

We close this section with some remarks on current conservation.
As noted previously,
for elastic electron deuteron scattering the current operator as
defined in the Bethe-Salpeter formalism is
conserved, provided the kernel of the Bethe-Salpeter equation
satisfies a local property.
 One of the crucial ingredients in showing this is that the
$\gamma NN$ vertex satisfies the Ward-Takahashi identity
\begin{equation}
\label{Ward}
q \cdot \Gamma^{(i)}  =  F_1^{(i)} ( {S^{(i)}}^{-1}(p^{(i)},P')-
{S^{(i)}}^{-1}(p,P) ).
\end{equation}
The analysis can readily be extended to the case of inelastic
electron scattering.
Note that in view of Eq.~(\ref{bsin}) the scattering wave
function  Eq.~(\ref{scat}) satisfies the homogeneous equation
\begin{equation}
\label{bsinhom3}
\tilde{\Psi}_{np}(p_f,p';P')S_2^{-1}(p',P') =  \frac{-i}{4 \pi^3}
\int d^4 k \tilde{\Psi}_{np}(p_f,k;P') V(k,p').
\end{equation}
Substituting the Ward-Takahashi identity in Eq.~(\ref{bsia2}) and
using Eqs.~(\ref{bshom}) and (\ref{bsinhom3}) leads in the same way
as in the elastic case to
\begin{equation}
\label{dd1}
q\cdot J^{IA}  = \frac{ 2e}{ (2\pi)^7 M_D} \sum_{i=1}^2
\int d^4 k \int d^4 p \tilde{\Psi}_{np}(k,P')  \left[
V(k,p^{i}),F_1^{(i)} \right] S_2(p,P) \Phi_D(p,P).
\end{equation}
The main difference with the case of elastic electron scattering is
that the final two-nucleon state is different from the initial state
and that the e.m. nucleon form factors in the
commutator also contain in this case in addition to the isoscalar also
an isovector part. If
the final state is an $I = 0$ state the form factors are isoscalar
and the commutator vanishes as for elastic scattering. In breakup
reactions however also the $I = 1$ channel is present in the
final state, yielding a non vanishing isovector contribution.
Consequently  MEC contributions should be added to Eq.~(\ref{dd1})
to make it divergenceless. These will be considered in
detail within the quasipotential approach in a forthcoming paper.

\section{ELASTIC SCATTERING IN ET APPROXIMATION}

The elastic deuteron current has precisely the same form as
Eq.~(\ref{loopqp2}) except that ${\tilde \Phi} \rightarrow
{\tilde \Phi_D}$.
It is worth noting that the deuteron current for elastic
scattering is  conserved
in the ET approximation at the level of positive energy states.
This can be seen as follows.
Keeping only the positive energy state contributions
we get from Eq.~(\ref{loopqp2})
 \begin{eqnarray}\label{qdj3}
 q \cdot J^{IA}
  = &&  \frac{-e}{64 \pi^{5}M_D}\int d^{3}k \int d^3 p
 \tilde{\Phi}_D^{BSLT}(\hat{k};P')
 S^{BSLT}_2(\hat{k},P')
 \nonumber \\ &&
 \times \sum_{i=1}^2\left[V(\hat{k},\hat{p}^{i} ),
F^{(i)}_1 \right] S^{BSLT}_2(\hat{p},P) \Phi_D^{BSLT}(\hat{p};P),
 \end{eqnarray}
since $1/i\pi\int d p_0 S_2(p,P)$ is equal to
$S_2^{BSLT}(\frac{1}{2}\hat{p},P)$ for positive energy states.
Because now only the isoscalar parts of the form factors
$F_1^{(i)}$ contribute to the current matrix elements for
elastic scattering and these parts commute with the potential V, we
may indeed conclude that Eq.~(\ref{qdj3}) is divergenceless.

Since the normalization condition of the deuteron wave function
is related to current conservation, we should expect the correct
normalization only for positive energy states.
The normalization condition for the BSLT vertex function is
given by
\begin{equation}\label{nBSLT}
P_{\mu} \delta_{M,M'}=\frac{1}{(2\pi)^3} \int d^3k
 \tilde{\Phi}^{(M)}_D({k};P)  \left( \frac{\partial}{\partial
P_\mu} S_2^{BSLT}(k,P) \right)_{_{ P^2 = M^2_D }}
\Phi^{(M')}_{D}({k};P).
\end{equation}
This can readily be verified to hold. Using the identity
\begin{equation}\label{propid}
-\frac{\partial}{\partial P_\mu} S_2(p,P) = \frac{1}{2} [ S^{(1)}(p,P)
\gamma_\mu^{(1)} S_2(p,P)
+  S^{(2)}(p,P) \gamma_\mu^{(2)} S_2(p,P) ],
\end{equation}
we get for the ET current matrix element Eq.~(\ref{loopqp2})
in the limit $ q \rightarrow 0$
\begin{equation}\label{normqp}
J_\mu^{IA} =
 \frac{-2i e}{(2\pi)^4 M_D}\int d^4 p \tilde{\Phi}_D^{BSLT} (\hat{p},P)
\frac{\partial}{\partial P_\mu} S_2(p,P)
\Phi^{BSLT}_D(\hat{p},P),
\end{equation}
which is indeed identical to Eq.~(\ref{nBSLT})
provided that only the positive energy states are kept.
It should be noted that current conservation for the more
general case can in principle be restored by
adding an effective two-body current to Eq.~(\ref{qdj3}) to cancel the
four divergence of the IA current matrix element. Because the
violation is at the level of the negative energy state
contributions, the correction is of relativistic order.

The deuteron current (\ref{loopqp2}) as discussed in the previous
section is obviously
different from the version studied in \cite{hutj}, where
the BSLT two-body propagators were used for both the
initial and final state propagators.
Here we have kept the propagator structure and the
$p_0$-integration is performed only over the propagators, since
we have implicitly assumed that the $p_0$-dependence may be
neglected in the vertex functions. We now compare the elastic electron
scattering results as obtained with the two approaches.
The calculations  in the ET approach proceed in exactly the same way
as in \cite{hutj,ZuTjGr}.  Using the helicity framework
(see Appendix A) we get for the current matrix element
 \begin{eqnarray}\label{prgpet}
 <P',M'|J_\mu^{IA}|P,M> &  = &
 \frac{2 i e}{(2\pi)^{4} M_D}
  \int dk_{0} \int dk k^{2} \int d \Omega_k
  \nonumber \\
 & & \times
 \sum_{n n'}  \sum_{m,m'} \sum_{\tilde{\rho}, \tilde{\rho}'}
 \tilde{\phi}_{n',m'}^{BSLT}({\bf k}')S^{(1)}_{\rho' \tilde{\rho}'} (k')
 \tilde{\Gamma}^{(1)}_{\tilde{m}',\tilde{m}}
 S_{2_{\tilde{\rho}\rho}}(k)
 \phi_{n,m}^{BSLT}({\bf k}),
  \nonumber \\
 \end{eqnarray}
where we use the notation $m = \{\lambda_1, \lambda_2, \rho\}$
$\tilde{m} = \{\lambda_1, \lambda_2, \tilde{\rho}\}$, with
$\lambda_i$ being the helicity of the i-th nucleon and $\rho$
the quantum numbers of the rho-spin of the two-nucleon state.
The angular quantum numbers are characterized by $n = \{J,M,L,S \}$.

The current matrix elements are evaluated in the
Breit frame. Both initial and final states are boosted to their
c.m. frame. On the $\rho$-spin
basis $|++>$, $|-->$, $|e>$ and $|o>$ both propagators in
Eq.~(\ref{prgpet}) can be written in matrix form
\begin{equation}\label{mat}
 \left( \begin{array}{cccc}
   S_{++} & 0 & 0 & 0 \\
   0 & S_{--} & 0 & 0 \\
   0 & 0 & S_{ee} & S_{eo} \\
   0 & 0 & S_{oe} & S_{oo}
  \end{array}  \right).
\end{equation}
For the single nucleon propagator we have
$ S_{++}  =  S^{(1)}_+(k')$,
$S_{--}  =  S^{(1)}_-(k')$,
$S_{ee}  = S_{oo} =  (S^{(1)}_+(k')+S^{(1)}_-(k'))/2$ and
$S_{eo}  = S_{oe}=  (S^{(1)}_+(k')-S^{(1)}_-(k'))/2$.
The two-nucleon  Green function $ S_{2_{\tilde{\rho}\rho}}(k)$ can also
be written in the form (\ref{mat}) with
$S_{++}  =  S_{2_{++}}(k)$,
$S_{--}  =  S_{2_{--}}(k)$,
$S_{ee}  =  S_{oo} = (S_{2_{+-}}(k)+S_{2_{-+}}(k))/2$ and
$S_{eo}  = S_{oe} =  (S_{2_{+-}}(k)-S_{2_{-+}}(k))/2$.
For the explicit form of these propagators we refer to Appendix
A.  We note
that the matrices are no longer diagonal as is the case for the
choice made in \cite{hutj} where the BSLT propagators are used.

In Fig.~\ref{elcq} are shown the results for
the electric form factors $F_C$ and $F_Q$. The dashed line and the
dotted-dashed line give respectively the results when all components or
only positive energy states in
the deuteron vertex function are included.
We see that only at high momentum transfer the negative energy
components contribute noticeably to the form factors. For comparison
also are shown the results for the BSLT reduction as discussed
in \cite{hutj}.
In both calculations the H\"{o}hler et al. \cite{Ho}
$\gamma NN$ form factors have been used.
We see that the dip in
the charge form factor $F_C$ is sensitive to the model used. At low
momentum transfer the models give results in good agreement with
each other. Only at high $q^2$ the results differ. In Fig.~\ref{elmag}
are displayed the results for the magnetic form factor $F_M$.
Negative energy
components in the wave function yield large contributions to the form
factor in the dip region. When these components are included, the
position of the dip is shifted to higher momentum transfer. The two
quasipotential reductions give different results when only positive
energy states are taken into account. However, inclusion of the
negative energy states results in a good agreement between the two
approximations.
The results for the electric form factor A and the magnetic form
factor B are shown in Fig.~\ref{elAB} using the
ET approximation.
In the calculations of the ($\rho \pi \gamma$ and $\omega
\varepsilon \gamma$) MEC the BSLT propagator has been used
instead of the ET one.
They only differ (see Appendix \ref{helicity}) for negative energy
states. Calculations in
\noindent
various kinematic situations show that
such a replacement doesnot affect the MEC results noticeably.
The complete ET calculations of A yield essentially the same results as
those with the BSLT approximation, except for some small
differences at high momentum transfer.
For the magnetic form factor the results deviate only in the dip
region. The dip position in the ET
approximation is at too low momentum transfer as compared to the data.
Since the dip region is sensitive to the MEC coupling constants,
no specific conclusions can be drawn from this. In particular,
adapting the unknown
$\omega \varepsilon \gamma$ coupling constant one can readily
reproduce the experimental data.
Also the tensor polarization $t_{20}$ are similar in both calculations.
They agree with the experimental data
(Fig.~\ref{elt20}). At $q > 4 \mbox{fm}^{-1}$ the predictions of
the ET approximation are somewhat above the results of the
BSLT approximation, which is due to a lower value of the dip
in the charge form factor.

\section{ELECTROMAGNETIC BREAKUP OF THE DEUTERON }

We now turn to describe the procedure to calculate the structure
functions for inelastic electron scattering.
There are two terms which are calculated separately. One is the
PWIA and the other one is the FSI contribution.
To determine the current matrix element we
need to know the isospin structure.
For the free final state we have  both isospin $I = 0$ or $I = 1$,
whereas the deuteron is an isospin $I = 0$ state.
Clearly, the isospin has to be chosen for a given total spin $S'$
of the two-nucleon system in accordance with the final state being
totally antisymmetric.
Since the outgoing nucleons are on-mass shell  we have the conditions
$P_0' = E_{\frac{1}{2} {\bf P}'+{\bf p}_f}
+E_{\frac{1}{2} {\bf P}'-{\bf p}_f}$ and $p_f \cdot P' = 0$.
In the following we consider the e.m. current matrix element with
the final np pair in a total spin $(S',M'_S)$ state.

\subsection{The PWIA contribution}

The simplest contribution to the deuteron current is the coupling
of the photon to one of the two nucleons in the deuteron without
FSI between the two outgoing nucleons.
In Fig.~\ref{pwiad} all possible contributions are depicted.
Both initial and final states are antisymmetric
and therefore the contributions (a) and (b) are identical.
Let us now adopt the convention, where particle 1 in the final
state is assumed to be  the proton, carrying
momentum $\frac{1}{2} P' + p_f$.
To determine the structure functions (\ref{rlab}) we
calculate the e.m. current matrix elements in the c.m. frame of the
final state. In this frame we have $P_0'=2 E_{pf}$ and $p_{f0} =
0$.
The PWIA contribution can be written as
 \begin{eqnarray}\label{pwia1}
<P', p_f, S', M'_S  | J_\mu^{PWIA}&& |P,M>  =
\frac{ ie \sqrt{2}}{ 4  \pi M_D}  \sum_{i=1}^2
\nonumber \\ &&\times
  <P',p_f,S', M'_S| \Gamma_\mu^{(i)} S^{(i)}
 (p^{(i)}, P) \Phi^{(M)}_D(p^{(i)}, P),
 \end{eqnarray}
where
$p^{(1)} = p_f- \frac{1}{2} q$, and $p^{(2)} = p_f+\frac{1}{2} q$.
The state $  |P', p_f, S', M'_S  >$
describes the free np pair with
total spin $S'$ and its explicit form is given in Appendix
\ref{helicity}.

Since the deuteron vertex function is usually calculated in
its rest system, it has to be boosted to the c.m. system of the
final np pair. The Lorentz transformation which does this is
given in Eq.~(\ref{lortr}).
The corresponding one for spin-$\frac{1}{2}$ particles is given
by
\begin{equation}
\Lambda( {\cal L}) = \sqrt{\frac{E_{D} + M_{D}}{2 M_{D}}} \left[ 1 +
\gamma^{0} \gamma^{3}\frac{-q}{E_{D} + M_{D}} \right].
\end{equation}
Boosting the initial state to the
c.m. frame of the np pair, the deuteron current contribution
Eq.~(\ref{pwia1}) can be written in the form
 \begin{eqnarray}\label{pwia2}
<P', p_f, S', M'_S  | J_\mu^{PWIA}&& |P,M>  =
\frac{ ie \sqrt{2}}{ 4  \pi M_D}  \sum_{i=1}^2
\nonumber \\ && \times
 <P',p_f,S',M'_S| \tilde{\Gamma}_\mu^{(i)} S^{(i)}
 (k^{(i)}, P) \Phi^{(M)}_D(k^{(i)}, P),
\end{eqnarray}
where $\tilde{\Gamma}_\mu^{(i)} = \Gamma_\mu^{(i)} \Lambda({ \cal
L})$, $  \Lambda({ \cal L}) = \Lambda^{(1)}({ \cal L})
\Lambda^{(2)}({ \cal L})$ and $k^{(i)} = {\cal L} p^{(i)}$.

Introducing the deuteron
state components $\phi_{n,m} ({\bf p}, p_0)$ labeled
by $n = \{ J, M, L,
S\}$ and a combined helicity and $\rho$-spin label $m = \{
\lambda_1, \lambda_2, \rho \}$, the deuteron current Eq.~(\ref{pwia2})
can now be evaluated in the same way as has been done in the elastic
case.
For a more detailed discussion we refer to
\cite{hutj}. On the chosen basis the deuteron current is found
to be of the form
 \begin{eqnarray}\label{pwia3}
<P',p_f, &&S', M'_S| J_\mu^{PWIA}  |P,M>   =
 \frac{ ie \sqrt{2}}{ 4 \pi M_D}   \sum_{i=1}^2
 \nonumber \\ &&\times
   \sum_{m'} \sum_{\tilde{\rho} m,n}
 D^{S'}_{M_S' \lambda''}(\Omega_f)
 C^{\frac{1}{2} \frac{1}{2} S'}_{\lambda_1'' -\lambda_2'' \lambda''}
 \tilde{\Gamma}_{\mu, m', \tilde{m}}^{(i)}
  S^{(i)}_{\tilde{\rho} \rho}
 (k^{(i)}, P) \phi_{n,m}({\bf k}^{(i)}, k_0^{(i)}),
\end{eqnarray}
where $m' = \{\lambda_1', \lambda_2', ++\}$ and the explicit
form of the
one-particle propagator $S^{(i)}_{\tilde{\rho} \rho} $ is given in
Appendix \ref{helicity} and Eq.~(\ref{mat}). The momentum
$k^{(i)}$ is restricted by momentum conservation, since the
spectator particle is on-shell we get that $k^{(1)}_0 = E -
E_{\vec{k}^{(1)}}$ and $k^{(2)}_0 = -E + E_{\vec{k}^{(2)}}$.
In writing down Eq.~(\ref{pwia3}) we have
shifted a $\gamma_0$
 matrix to the vertex, so that the vertex operator is now given by
$\tilde{\Gamma}_\mu^{(1)} = \Gamma_\mu^{(1)} \Lambda({ \cal
L})\gamma_0^{(2)}$ and $\tilde{\Gamma}_\mu^{(2)} = \Gamma_\mu^{(2)}
\Lambda({ \cal  L})\gamma_0^{(1)}$ . The extra  $\gamma_0$ factor
is due to the use of
the closure relation for the helicity spinors. In
Eq.~(\ref{pwia3}) the $\varphi$-dependence can explicitly be evaluated.
For the spherical components $\Gamma_\mu$ (with $\mu=0,\pm
1$) of the e.m. current we find
that the total $\varphi$-dependence is given
by $\exp[i(M+\mu-M_S')\varphi]$.
With this $\varphi$-dependence of the current operator,
$R_L$ and $R_T$ don't
depend on $\varphi$ and is the $\varphi$-dependence of $R_{TL}$ and
$R_{TT}$ given respectively by $\cos \varphi$ and $\cos 2\varphi$.
The isospin structure contained in $\Gamma_\mu$ can readily be
evaluated. Since particle 1 is assumed to be the proton we get
$\Gamma_\mu^{(1)} \rightarrow F_p/\sqrt{2}$ and
$\Gamma_\mu^{(2)} \rightarrow F_n/\sqrt{2}$.
In the actual calculations it is convenient to treat the proton
and neutron contributions to the e.m. current separately. Denoting
the term where the photon interacts with the proton (neutron)
as $J^{(p)}_\mu$ $\left( J^{(n)}_\mu\right)$ we find with the
help of symmetry relations for the deuteron vertex functions
(see \cite{hutj})
\begin{equation}\label{neutprot}
J^{(n)}_\mu ({\bf p}_f) = (-1)^{L+S+S'} J^{(p)}_\mu (-{\bf p}_f).
\end{equation}
with all e.m. proton form factors in $J^{(p)}_\mu $ replaced by the
corresponding ones of the neutron.

The ET approximation for the PWIA contribution can be
obtained immediately from the above expressions, by simply
replacing the initial
deuteron bound state by the BSLT vertex function. The relative
energy variable in the propagator is prescribed by the condition
that the final state describes on-shell particles.

\subsection{Rescattering contribution}

When FSI is included in the IA contributions, the structure is
more complicated.
Starting from the full Bethe-Salpeter formalism the current matrix
elements has the form given in Eq.~(\ref{bsia2}). We have
 \begin{eqnarray}\label{fsi1}
<P',p_f, S',M'_S |&& J_\mu^{FSI} |P,M>
 =  \frac{e}{ (2 \pi)^4 M_D} \sum_{i=1}^2\int d^4p'
\nonumber \\ && \times
 {\tilde \phi}_{np}(p_f,p',P')
 S_2(p',P') \Gamma_\mu^{(i)}S^{(i)} (p^{(i)},P) \Phi^{(M)}_D
 (p^{(i)},P) .
 \end{eqnarray}
 Assuming again that particle 1 in the final state is the
 proton, we only have to take
the two graphs (a) and (b) into account as represented in Fig.~\ref{fsid}.
Using  the same analysis leading to Eq.~(\ref{pwia3}) we find
 \begin{eqnarray}\label{fsi2}
 \lefteqn{<P',p_f,S',M'_S | J_\mu^{FSI} |P,M>  =
 \frac{  e }{ (2\pi)^4  \sqrt{2} M_D}
\sum_{i=1}^2
\sum_{n',n}
 \sum_{m,m',m''} \sum_{\tilde{\rho}
 \tilde{\rho}'}
\int d^4 p'}
 \nonumber \\
 \times
 & & D^{S'}_{M_S' \lambda'}(\Omega_f)
 C^{\frac{1}{2} \frac{1}{2}
 S'}_{\lambda_1' -\lambda_2' \lambda'}
 {\tilde \phi}_{n';m',m''}(p_f,p')
 S_{2_{\rho'' \tilde{\rho}''}}\tilde{\Gamma}_{\mu, \tilde{m}''
 \tilde{m}}^{(i)}  S^{(i)}_{\tilde{\rho} \rho}
 (k^{(i)}, P) \phi_{n,m}({\bf k}^{(i)}, k_0^{(i)}) ,
 \end{eqnarray}
where the partial wave components ${\tilde \phi}_{n';m',m''}(p_f,p')$
correspond to the isospin dependent t-matrix elements with
total angular momentum
$n' = \{J', M'\}$ between the energy-helicity spin states
$m' = \{\lambda_1', \lambda_2',
++\}$ and $m'' = \{\lambda_1'', \lambda_2'', \rho''\}$ .
The isospin dependence of $\tilde{\Gamma }$ in Eq.~(\ref{fsi2})
is for an isoscalar
transition given by $F_S/\sqrt{2}$ and for an isovector transition
$(-1)^i F_V/\sqrt{2}$.

Replacing the integration variable $p'$ by $-p'$ in the integral
we readily see that both contributions are equal.
Moreover, collecting all the $\varphi-$dependence we find
$e^{i(M'-M_S')\varphi_f} e^{i(M+\mu - M') \varphi'}$,
so that the $\varphi'$-integration can be performed with the result
\begin{equation}
2 \pi \delta(M+\mu - M')e^{i(M+\mu - M_S') \varphi_f}.
\end{equation}
Notice that the same $\varphi_f$-dependence exists as for the
PWIA.
In both the initial and final state the relative
energy variable is set to zero, corresponding the BSLT
prescription.  Since the only
$p_0'$-dependence is in the nucleon propagators, the
$p_0'$-integration can be done analytical.
As a result we are left with a
two-dimensional integral which can be done numerically using standard
gaussian quadratures.  In so doing, one particular point has to
be taken care of.
At the point $E_{p'} = \frac{1}{2} M_{np}$ the particles
are on mass-shell and we have a singularity in the positive energy
propagator, which is of
the form $1/(E_{p'}-\frac{1}{2} M_{np})$. This
singularity can simply be dealt with using a standard
subtraction technique.

Having determined the e.m. current matrix elements in this way the
deuteron structure functions can be constructed using
Eq.~(\ref{Rk}). It should be noted that not all partial wave
matrix elements in Eq.~(\ref{fsi2}) need explicitly to be calculated.
Due to parity conservation we have
\begin{equation}
\label{symm1}
 <J'M'L'S'| J_\mu |P,M> =
(-1)^{J+J'+M+M'+\mu} <J'-M'L'S'| J_\mu |P,-M>.
\end{equation}
Besides this symmetry relation we have the selection rule
$M'=M+\mu$ for the above matrix element, being a consequence of
rotational symmetry.

\section{RELATIVISTIC ANALYSIS OF INELASTIC ELECTRON SCATTERING}

In the preceding sections we have developed a relativistic
framework to describe electrodisintegration of the deuteron.
The full relativistic form of the e.m. operator is employed and
at the same time the relativistic structure of the nuclear
dynamics is incorporated.
To see what kind of predictions our fully relativistic formulation
leads to in the case of inelastic electron scattering we study the
structure functions for two kinematic situations. One is the inclusive
experiment carried out by the Bates-Mit group and the other is
the exclusive analysis by the Nikhef-K group.
Since we also want to compare our predictions to those obtained
from nonrelativistic calculations, let us first discuss the e.m.
operator used in the Schr\"{o}dinger approach.

\subsection{Nonrelativistic limit}

In most of the studies of electron-deuteron scattering it is
implicitly assumed that to a good approximation the deuteron can
be described by a boundstate wave function satisfying a
Schr\"{o}dinger equation. In addition some effective nonrelativistic
(NR) form is used for the e.m. operator.
Various versions have been proposed in the
literature based on taking the NR limit of the current operator.
The e.m. operator is expanded in
$q/M$, keeping only the first order term in the expansion and
neglecting recoil effects. Because NR wave functions are used,
boost effects are in general also neglected. The resulting
effective e.m. operators have all in common to be the same
in leading order, but to differ in relativistic order $(q/M)^2$
Since we have included in our study the full relativistic structure
of the e.m. operator, it is obviously of interest to see how these
relativistic corrections look like.
In the calculations the vertex is defined in the cm-frame of the final
state. Because the helicities in this frame and the cm-frame of
the initial deuteron state are not equal, we have to replace the
boost operators by Wigner rotations. A detailed discussion is
given in Ref. \cite{hutj}.

In the NR limit all particles are taken to be on-shell and
the negative energy state components are neglected, i.e.
since the nucleons are assumed to behave nonrelativistically,
we may assume that the nucleons can be described by the positive
energy spinors
\begin{eqnarray}
 u_{\lambda}^{(1)}({\bf p})  &=&  N_{p}\left[ \begin{array}{c} 1 \\
                            \frac{{\bf \sigma}.{\bf  p}}{E_{p}+M}
                                \end{array} \right]
                \chi_{\lambda},
\nonumber \\
 u_{\lambda}^{(2)}({\bf p})  &=&  N_{p}\left[ \begin{array}{c} 1 \\
                            \frac{-{\bf \sigma}.{\bf  p}}{E_{p}+M}
                                \end{array} \right]
                \chi_{-\lambda} ,
\end{eqnarray}
where $N_p = \sqrt{E_p + M_N/2 E_p}$ and
$\chi_{\lambda}$ are the Pauli spinors with helicity $\lambda$.
Therefore the current matrix element to study in this
approximation is given by
\begin{equation}\label{complete}
J_\mu =  \overline{u}_{\lambda'}({\bf p}') \left(F_1 \gamma_\mu +
\frac{i}{2 M_N} F_2 \sigma_{\mu \nu} q^\nu\right)u_{\lambda}({\bf p}).
\end{equation}
It can immediately be reduced to a Pauli form. We get
\begin{eqnarray}\label{second}
J_0  & =  & F_1 - \frac{1}{8 M_N^2} (F_1+2 F_2)({\bf q}^2 +i {\bf \sigma}
\cdot [{\bf K} \times {\bf q}]) + O ( \frac{q^4}{M_N^4} ) \nonumber
\\
{\bf J} & = & \frac{F_1}{2M_N} \left[ ({\bf K}+i [{\bf \sigma} \times
{\bf q}]) (1- \frac{1}{8 M^2}\{{\bf K}^2 +{\bf q}^2 \}) - \frac{1}{8
M^2}({\bf q}+ i [{\bf \sigma} \times  {\bf K}]) ({\bf K} \cdot {\bf q})
\right] \nonumber \\
&& +  \frac{F_2}{2M_N}\left[ i [ {\bf \sigma} \times {\bf q}](1- \frac{1}{8
M^2}{\bf q}^2) -\frac{\omega}{2 M_N} ({\bf q}+ i [ {\bf \sigma}
\times  {\bf K} ] )
 \right.
 \nonumber \\ &&
 \left.
+ \frac{1}{8 M^2}\left( [{\bf K} \times {\bf q}] \times
{\bf q} - i ({\bf \sigma} \cdot {\bf K}) [{\bf K} \times {\bf q}]\right)
\right] + O ( \frac{q^4}{M_N^4} ),
\nonumber \\
\end{eqnarray}
\noindent
with ${\bf K} = {\bf p} + {\bf p}'$ and where we have kept all second
order terms.

In the usual NR limit only the first order terms are kept in the
charge and current operator.
In so doing, we get the operator as used by
Mathiot \cite{Ma84} and Sommer \cite{So}
\begin{equation}\label{maso}
J_0  =  F_1,\;
{\bf J}  =  \frac{1}{2M_N} F_1 {\bf K} - \frac{i}{2M_N}G_M
({\bf q} \times {\bf \sigma}).
\end{equation}
Note that the correction terms in Eq.~(\ref{second}) are a rather
complicated function of both $q$ and $K$. Except for the region
of low $q$ and small recoil corrections
there is no a priori reason to expect them to be small.

Another form which is  often used, is found by  starting
from the alternative form of the
relativistic current operator in terms of Sachs form factors
\begin{equation}\label{Sachs}
G_E  =  F_1 + \frac{q^2}{4 M_N^2} F_2,\;
G_M  =  F_1 + F_2.
\end{equation}
Taking the NR limit in this case yields the e.m. operator used
for example by Leidemann and Arenh\"{o}vel  \cite{LeAr}
\begin{equation}\label{lear}
J_0  =  G_E,
{\bf J} = \frac{1}{2M_N} G_E {\bf K} - \frac{i}{2M_N}G_M
({\bf q} \times {\bf \sigma}).
\end{equation}
The only difference between the two approximations is the use
of $F_1$ or $G_E$ in the charge and current operator. From
Eq.~(\ref{Sachs}) we see that the difference is of order $q^2/M_N^2$,
being of relativistic origin.
Considering Eq.~(\ref{second}) we see that the charge operator can
be rewritten up to order $q^4$ as
\begin{equation}\label{joge}
J_0  =  G_E(1 - \frac{{\bf q}^2}{8 M_N^2} )
- \frac{1}{8 M_N^2}(2 G_M- G_E)
i {\bf \sigma} \cdot ({\bf K} \times {\bf q}).
\end{equation}
Momentum conservation at the vertex gives ${\bf K} =
{\bf p}+{\bf p}' = 2 {\bf p} +{\bf q}$. Hence for small values of the
missing momentum, i.e. $p << q$, the second term in Eq.~(\ref{joge})
can be neglected. For values of $q$ up to a few
hundred MeV it suggests that the choice of $G_E$ is indeed
reasonable for the charge operator.
The procedure as followed in Eq.~(\ref{joge}) can not be used for
the current operator, because of the lack of a small parameter.
However, we can show that the isovector part up to order $q^4$ is
given by
\begin{equation}\label{jige}
- \frac{i}{2 M_N} G_M (1 - \frac{3 {\bf q}^2}{8 M_N^2}) ({\bf q} \times
\sigma),
\end{equation}
which shows that the choice of $G_M$ in Eq.~(\ref{maso}) and
Eq.~(\ref{lear}) is reasonable for small $q$.
It should be noted that in some calculations of Ref.
\cite{LeAr} the factor $C_{np}=\frac{M^2_N}{E_p E_n}$ has been added to
the cross section, whereas this term is already included in
Eq.~(\ref{second}). This correction is clearly of relativistic order.

Besides the additional terms neglected for a  given
choice of the NR e.m. operator, other relativistic effects like
contributions from the boost transformation and the negative
energy state components of the deuteron and scattering wave function,
can play an important role.
In Fig. \ref{relfrac} are shown the relative  contributions
from these corrections to the response functions for an
exclusive experiment with missing momentum $p_m$ = 100 MeV/c
and $q$ in the range of 300 MeV/c to 1500 MeV/c. In addition
we used $\theta_p = 120^o$ and $\theta_e = 60^o$ and took the
kinematics in plane. In this figure we are comparing the
different approximations to the complete relativistic
calculation with FSI included, but without negative energy state
contributions.
The effect of including the negative energy state contributions
is given by the double dot-dashed curve. As a second
approximation we may drop the boost operators in
the current operator (solid line). This is similar to the NR
approximations except that we keep the complete current operator
Eq.~(\ref{complete}). If only the
PWIA contribution is considered all response functions show
identical dependence on the boost operators.
The contributions from the boost to $R_L,
R_T$ and $R_{TL}$ become more important at higher $q$ values and
can be as large as 40 $\%$ around $q $ = 1500
MeV/c, whereas in the case of  $R_{TT}$
inclusion of the FSI substantially enlarges this dependence and
the boost effects are large even for low momentum transfer.
 The negative energy state contributions increase
at higher q and are of the order of $5 \%$ around q = 300 MeV/c.
Again a different behavior is found for $R_{TT}$, an effect of
around $30 \%$ is seen for the whole kinematic region.
An important conclusion is that the negative contributions
cancel to a large extent the boost effects. Similar as for
elastic electron deuteron scattering
a consistent relativistic treatment is also important in this case.

Next we may consider the effect of neglecting the FSI
contributions. This yields the long dashed curve in Fig. \ref{relfrac}.
As expected we see an increase in the FSI contribution at lower q,
since q is directly related to the energy of the np-pair in the
final state.
Also it is of some interest to see the sensitivity on the choice of the
relativistic two-nucleon propagator on the response functions.
The dashed line depicts the relative difference between the BSLT
and ET approximation, including FSI. Except for $R_{TT}$,
the difference is less than $5 \%$ and increases for higher values of q,
whereas for $R_{TT}$ it decreases with q. This dependence on the
propagator structure only exists if FSI is included.

Finally, also plotted in Fig. \ref{relfrac} are the predictions with the
e.m. operators, given by Eq.~(\ref{maso}) and Eq.~(\ref{lear})
with  the $F_1$ and  $G_E$ form factors. The results are respectively
given by the dotted and the dot-dashed lines. As expected we see
increasing deviations with increasing q.
For $R_L$ the NR $G_E$ operator yields clearly the closest resemblance
to the calculation using the complete current operator. The
latter is in accordance  with our discussion of Eq.~(\ref{joge}) that
the use of $G_E$ in the charge operator should be reliable in
the quasielastic region.
Beause of isovector dominance in the
contribution of $R_T$ both NR operators give the same results
and deviate only a few percent of the results with the exact
current operator, what is in accordance with Eq.~(\ref{jige}).
Finally we see from the figure that
the deviations are very large for the response functions $R_{TL}$
and $R_{TT}$ even at comparatively low momentum transfer.
These deviations are essentially due the interference structure of
the response functions, therefore probing those parts of the current
operators not described well by the NR operators.

Not shown in these figures are the results with the current
operator of Eq. (\ref{second}) where we have kept all second
order terms. Similar higher order corrections have also been
considered recently~\cite{Mosconi,Ar92}
For $R_L$ and $R_T$ although the results
agree reasonably well with the full calculations at moderate
momentum transfer they are found to be substantially different
at higher momenta. In general at high momentum transfers the
deviations are such that the complete e.m. operator including
boost effects is needed for a reliable analysis of the response
functions. For $R_{TT}$ and $R_{TL}$ calculations
with this second order operator show it gives comparable results
as with the complete operator. Deviations are within a few
percent.
At these low missing momenta no large model dependence of the
wave functions is expected. The deuteron wave function is mainly
S-wave in this region. We have tested the aspect of the non
relativistic reduction of the propagators. With this approximation we
find effects smaller than 0.5 $\%$.

To get some insight about sensitivities on FSI and negative energy
states in a larger part of the phase space in Figs. \ref{theta_fsi}
and \ref{theta_neg}
contour plots are shown where we have varied in addition $\theta_p$. For
the relative contribution of the FSI,  shown in Fig.
\ref{theta_fsi}, we find very similar effects in the various
kinematic regions for $R_L$ and $R_T$.
{}From these figures we see that both contributions to in
particular the $R_{TT}$ response function can be substantial.

\subsection{Structure functions of the exclusive reaction}

As an example we  discuss the exclusive
experiments carried out at Nikhef-K~\cite{mvds,vdst}.
In these experiments in addition to the measured final electron also
the outcoming proton is detected. As a result the cross section is
described by the four response functions $R_L, R_T, R_{TL}$ and
$R_{TT}$ given in  Eq.~(\ref{Rk}). $R_{TL}$ and $R_{TT}$ have an out
of plane dependence of respectively $cos \phi$ and $cos 2 \phi$.
Consequently $R_{TL}$ can be separated by carrying out two measurements
at $\phi = 0$ and $ \phi = \pi$, such that
${\bf p} \cdot {\bf q}$ is constant. The response function $R_{TL}$ is
interesting because through its interference structure
it is sensitive to the isovector part of the charge operator
and the isoscalar part of the current operator.  By measuring
only $R_L$
and $R_T$ no information is obtained about these terms because of the
dominance of the isoscalar charge and isovector current operators.
The  experiments were performed with a missing momentum $p_m$ in the
range of 40 to 180 MeV/c. In this kinematic region the FSI is
expected to give only small contributions due to the low missing
momentum and therefore the high relative momenta of the outcoming
nucleon pair. Because of the dominance of the PWIA contribution
the chosen kinematics  is very suitable to examine the
underlying structure of the one-body current operator.

In Fig. \ref{nikexpr} the results of
the Nikhef experiments together with our predictions are shown.
In all these calculations the OBE model has been used to
construct the continuum and deuteron vertex functions.
In these figures PWIA corresponds to our relativistic PWIA
calculation and the curve denoted by REL(-) contains in addition the FSI
(both positive and negative energy states).
For both $R_L$ and $R_T$ the FSI is small and increases with
increasing missing momentum.
Also are plotted the predictions obtained using a nonrelativistic
description with e.m. operators, given by Eq.~(\ref{maso})
and Eq.~(\ref{lear}) with  $F_1$ and  $G_E$ form factors.
The results show clearly that
the current operator with $F_1$ is too high as compared to the
experimental data for $R_L$ and the $G_E$ description gives a much better
agreement with both the data and the relativistic calculation.
The results for $R_{TL}$  and $R_{TT}$ are remarkable. Although
we are at relatively low momentum transfer, relativistic effects
are substantial.
We find that our relativistic calculations leads to
a good description of the $R_{TL}$ data, whereas
both NR predictions are clearly too low.  In this case
the NR $F_1$ operator is closer to the relativistic prediction.
Similar conclusions can be drawn for the out of plane
structure function $R_{TT}$.
It is worth noting that for this
case the FSI effect is considerably larger.

{}From these results we may conclude that in the kinematic
region considered the most important relativistic effects arise
from the higher order corrections to the e.m. operators
employed in the usual NR analysis.
In general the NR descriptions are clearly not adequate and
they can represent a poor approximation to the full
relativistic operator.

\subsection{Inclusive scattering in the quasielastic region}

We now turn to describe our calculations for the inclusive
reactions in the quasielastic region. These experiments were done
at Mit-Bates for
momentum transfers $q$ in the range of $300-500$
MeV/c \cite{Dytman}. The quasielastic peak is defined by
\begin{eqnarray}\label{qe-peak}
M_D + \omega = \sqrt{M_N^2 + q^2} + M_N.
\end{eqnarray}
In the considered experiments a longitudinal-transverse
separation has been carried out. The four momentum transfer is
such that Eq.~(\ref{qe-peak}) to a good approximation reduces to
$\omega = q^2/2M_N $.
For these values of $\omega$ and $q$ the
electron is essentially probing the nucleons
with missing momentum near zero.

In Figs. \ref{respf1}-\ref{respf3} are shown the
various corrections to the longitudinal and transverse response
functions at momentum transfer q = 300, 400 and 500 MeV/c.
The full relativistic expression of the electromagnetic
vertex is used. Inclusion of the boost operator shows
only a very small contribution.
Considering $R_L$ we see that the PWIA describes to a good
approximation the response, except at lower momentum transfer.
The Born contribution can
essentially be neglected, because $F_1^n$ is very small.
On the other hand the FSI is seen to be important for $R_T$ over
the whole range of considered momentum transfer.
Moreover, since the magnetic form factor of the nucleon is comparable in
magnitude to $F_2^n$ the Born term also gives a substantial
contribution to $R_T$. For larger $\omega$ the delta-degree of
freedom is seen to come up. This contribution and the other MEC's
will be discussed in the same relativistic framework in a next
paper.

Including  only respectively the Born, the J = 0 and 1 NN
scattering amplitude in the two response functions we see that
most of the FSI has been accounted for .
The main FSI contribution to $R_L$
is from the $J = 1 \mbox{ }({}^1P_1, {}^3S_1- {}^3D_1,{}^3P_1)$
amplitudes. Especially at low $\omega$ this contribution
can be significant  due to the decreasing relative
momentum of the outcoming nucleon pair. Similarly
as for the three-nucleon systems~\cite{metj}
the response increases at low
$\omega$, whereas it decreases at the quasielastic peak due to
the FSI contribution.
The PWIA result for $R_T$ is clearly below the data.
In particular, the $ J = 0 \mbox{ }({}^1S_o, {}^3P_o)$ FSI
contribution is only of some importance for $R_T$.
Due to the FSI contributions our overall predictions are well in
accordance with the experimental data. Although substantially
smaller than in the
the trinucleon system some strength tends to be missing
in the transverse response function at lower energy transfer.
We also considered the negative energy state contributions.
Exactly as we have found for exclusive scattering this contribution
seems to be vanishing at the quasielastic peak.  Furthermore
it reduces the strength at low $\omega$, whereas it tends to
increase slightly the response functions past the quasielastic peak.

The overall agreement with the experimental data is gratifying.
Since at these relatively low momentum transfers the existing
parameterizations of the e.m. nucleon form factors are very
similar,
less dependence is found on the choice of these  nucleon form factors.
Using for example the
Iachello et al~\cite{Iach} form factors changes of only
at most 1 percent were found.

\section{SUMMARY AND OUTLOOK}

We have presented in this paper a framework to study the e.m.
deuteron current based on a relativistic quasipotential
approach. The nuclear dynamics itself is described in terms of a
symmetric choice for the relative energy variable. Assuming that
the two-nucleon vertex functions  depend smoothly on this
variable, the current matrix element is determined by explicitly
integrating out the relative energy variable in the e.m. vertex
loop. This equal time approximation has the advantage that it can
be used to describe both  elastic and inelastic electron scattering
and that it in principle can be extended in a systematic way.
As a direct application the elastic e.m. properties of the deuteron
are studied using this formalism.
The differences between this approach and the one used in
Ref.~\cite{hutj} for elastic electron deuteron scattering are
the treatment of the initial and final states and the
nucleon-propagator structure of the one-body current. We have
shown that for elastic scattering these two approaches give
compatible results and differences are found to be minor.

The approach discussed here has clearly the advantage that
besides elastic electron scattering case, it can also be applied
to the deuteron breakup in electro- and photo-disintegration
processes.
Explicit expressions are written down for the deuteron current in the
case of inelastic electron scattering. The helicity formalism is
extensively used to relate the scattering wave functions of the
relativistic one boson exchange model to the current matrix
element. Since there are also isovector current
contributions in this case, mesonic exchange currents are in
general needed to satisfy gauge invariance. These currents can
also be constructed in this ET approximation and will be studied
in a subsequent paper.

Various different forms of one-body operators exist in the
literature  to describe the e.m. processes in a nonrelativistic
framework. These are derived from relativistic operators which
are on-shell equivalent, but lead to different operators in the
NR limit.  We have compared the NR calculations with
our relativistic description of the electro disintegration of
the deuteron. The important difference is that we also take in
our approach the higher order terms in  $q/M$ into account.
It should be noted that we have assumed a specific on shell form
of the e.m. nucleon form factor and not considered its off-shell
dependence. For this a detailed dynamical model of the nucleon
is needed.

One interesting experiment to study the difference between a
relativistic and nonrelativistic analysis is
the exclusive experiment performed at Nikhef-K, in which
the three response functions $R_L, R_T$ and $R_{TL}$ could be
separated. Due to the small missing momentum in these
experiments the FSI-contribution is very small and the response
functions are in a good approximation described by the one-body
current (PWIA).
In our analysis we have found that both relativistic and
nonrelativistic theories
give a good description of $R_T$. Differences show up for $R_L$,
the NR-theory using the $G_E$ form factor gives a good
description, while the $F_1$ current operator is systematic to
high. Interesting results were found for the interference
structure function $R_{TL}$. Both NR-approaches failed to
describe the data. Including the next order in $q/M$ to the
one-body current operator showed a much better agreement.
Other relativistic effects have also been considered. These are
due to boost effects and the presence of negative energy states.
In the kinematics studied these contributions were found to be
small. However, it is expected that these corrections will be
substantial at higher momentum and energy transfer as can be seen
in the case of elastic electron scattering.
We have tested this assumption by considering the range of $q = 300$
MeV/c to $1500$ MeV/c. Effects of $40 \%$ were found.
The second experiment we discussed in our relativistic approach
is the inclusive experiment done at Bates. In this experiment
much larger effects due to the Born term and FSI are found. A
good agreement with the data is found, although at larger
momentum transfer $q$ the longitudinal response function is
slightly overestimated. This may however readily be explained in
terms of model dependence on the deuteron structure.
For these experiments the NR-description
using $G_E$ gives similar results as the relativistic calculation.

\appendix
\section{ HELICITY BASIS}
\label{helicity}

In this Appendix we summarize the helicity basis used in the
partial wave decomposition of the NN amplitude. For more details
we refer to Refs.~\cite{FlTj,Ku,TjWa}.
Following Kubis \cite{Ku} we use for the positive and negative energy
spinors
\begin{eqnarray}
\label{ku}
  u_{\lambda}^{(+)}({\bf p}) &=& N_{p}\left[ \begin{array}{c} 1 \\
                         \frac{2\lambda p}{E_{p}+M} \end{array} \right]
                \chi_{\lambda} (\theta,\varphi),
\nonumber \\
  u_{\lambda}^{(-)}({\bf p}) &=& N_{p}\left[ \begin{array}{c}
                           \frac{-2\lambda p}{E_{p}+M}
                             \\ 1 \end{array} \right]
                \chi_{\lambda}(\theta,\varphi),
\end{eqnarray}
where $\chi_{\lambda}(\theta,\varphi)$  are the two-component Pauli
spinors, which are given by
\begin{eqnarray}\label{paulirot}
 \chi_{\frac{1}{2}}(\theta,\varphi)  &=&  \left( \begin{array}{c}
cos \theta /2 \\
     e^{i\varphi}\sin \theta /2 \end{array} \right),
\nonumber \\
 \chi_{-\frac{1}{2}}(\theta,\varphi)  &=&  \left( \begin{array}{c}
                                -e^{-i\varphi}\sin \theta /2  \\
                                 \cos \theta /2 \end{array} \right).
\end{eqnarray}
These spinors are normalized as
\begin{equation}
u_\lambda^{(\rho)^\dagger} ( {\bf p}) u_{\lambda '}^{(\rho ')} ({\bf p})
= \delta_{\rho \rho '} \delta_{\lambda \lambda '},
\end{equation}
where $\rho = \pm$ and the normalization is given
by $N_p = \sqrt{(E_p+M_N)/2 E_p}$.
Let us consider the projection operators
\begin{equation}
\Lambda_\rho ( {\bf p}) = \sum_\lambda
u_{\lambda}^{(\rho)}({\bf p})\overline{u}_{\lambda}^{(\rho)}({\bf p}).
\end{equation}
We may rewrite the single nucleon propagator
\begin{equation}
S (p) = (
{p \settowidth{\letw}{$p$}
 \hspace{-0.4\letw}
 \makebox[0cm]{/}
 \hspace{0.4\letw}}
- M_N + i \varepsilon)^{-1} = \frac{
{p \settowidth{\letw}{$p$}
 \hspace{-0.4\letw}
 \makebox[0cm]{/}
 \hspace{0.4\letw}}
+
M_N}{p^2 - M_N^2 +i \varepsilon}.
\end{equation}
in terms of these projection operators. Using the relation
\begin{equation}
{p \settowidth{\letw}{$p$}
 \hspace{-0.4\letw}
 \makebox[0cm]{/}
 \hspace{0.4\letw}}
+M_N = (E_p + p_0) \Lambda_+( {\bf p})+ (p_0 - E_p )
\Lambda_-( {\bf p})
\end{equation}
we get for the propagator
\begin{equation}
S(p) = \frac{\Lambda_+( {\bf p})}{p_0-E_p + i \varepsilon}
+\frac{\Lambda_-( {\bf p})}{p_0+E_p + i \varepsilon}.
\end{equation}

We now turn to discuss briefly the partial wave representation
of the two-nucleon t-matrix.
When two-particle states are involved,
we take the convention of Jacob and Wick \cite{JaWi} where
particle 1 is described by the spinors $u_{\lambda}^{(\rho)}
({\bf p})$, while for particle 2 the spinors $u_{-\lambda}^{(\rho)}
(-{\bf p})$ are used.
The two-particle states built from these helicity spinors form a
complete basis $| {\bf p} \lambda_1 \lambda_2 \rho >$ (with
$\rho=(\rho_1,\rho_2)$) in Dirac space.
In this representation the t-matrix is given by
\begin{equation}\label{tmat1}
<{\bf p}_f \lambda_1' \lambda_2'  \rho'| \phi(p_f, p,P)  |{\bf p}
\lambda_1 \lambda_2  \rho>.
\end{equation}
Its angular dependence can be exhibited using the total angular
momentum states
\begin{equation}\label{angmom}
| J M \lambda_1 \lambda_2 > = \sqrt{\frac{2 J+1}{2}}
\int d \Omega_p
D^{J^*}_{M \lambda} ( \theta, \varphi) |p, \theta, \varphi,
\lambda_1, \lambda_2>.
\end{equation}
In view of rotational invariance of the t-matrix, Eq.~(\ref{tmat1})
can be written in the form
\begin{equation}\label{tmat2a}
t \equiv \sum_n \phi_{n;m',m}(p_f,p)
= \sum_{J,M} \frac{2J+1}{2}  D^{J^*}_{M \lambda'} ( \Omega_f)
\phi^J_{m',m} (p_f,p)  D^{J}_{M \lambda} ( \Omega) ,
\end{equation}
\noindent
where we have defined $ \phi^{J}_{m',m} (p_f,p) = <p_f J
\lambda_1' \lambda_2'  \rho'| \phi(p_f,p,P) |p J \lambda_1
\lambda_2  \rho> $ with $n = \{ J,M\}, m' = \{
\lambda_1', \lambda_2', \rho'\}$ and $m = \{  \lambda_1,
\lambda_2, \rho\}$. These angular momentum states can be related to
states labeled by the set $\{J M L S \}$.
We have
 \begin{eqnarray}
 t  =  \sum_{J,M} &&\sum_{L',S',L,S}
 \sqrt{\frac{2 L' +1}{2}} \sqrt{\frac{2 L +1}{2}}
 \nonumber \\ && \times
  D^{J^*}_{M \lambda'} (\Omega_f) C^{L' S' J}_{0 \lambda'
 \lambda'} D^{J^*}_{M \lambda} (\Omega_p) C^{L S J}_{0 \lambda
 \lambda} <J L' S' | \phi^J (p_f, p ,P) | J L S >.
 \end{eqnarray}
A  detailed analysis is given in the Refs.~\cite{FlTj,Ku,TjWa}.
The free np pair can be described by the two-particle
states $| {\bf p} \lambda_1 \lambda_2>$ with $\rho= (+,+)$.
These helicity states are
connected to the total spin $S$ states $ |{\bf p}, S, M_S>$ through the
relation
\begin{equation}
| {\bf p}, S, M_S> = \sum_{\lambda_1 \lambda_2}
D^{S^*}_{M_S
\lambda} (\Omega_p) C^{\frac{1}{2} \frac{1}{2} S}_{\lambda_1
-\lambda_2 \lambda} | {\bf p} \lambda_1 \lambda_2> .
\end{equation}

\section{ THE $k_0$ loop INTEGRATION IN THE E.M. CURRENT}
\label{k0int}

In  the current operator for the IA in the equal time
approximation, initial and final state are assumed to be independent
of the relative energy variable. As a result the only dependence on
this variable is in the nucleon propagators. Let us first
consider the inelastic case. The integral we
want to study is of the form
\begin{equation}\label{int1}
I = \int d k_0' S_2(k',P') \tilde{\Gamma}_\mu S^{(1)}(k,P)
\end{equation}
where $\tilde{\Gamma}_\mu$ is the e.m. operator including the
boost operators.  Both sets of
variables $(k', P')$ and $(k, P)$ are in their own c.m. frame, i.e.
$P' = ({\bf 0}, 2E')$ and $P = ({\bf 0}, 2E)$ where $2 E' = M_{np}$
and $ 2 E  = M_D$. The
Lorentz transformation ${\cal L }$ of the c.m. frame to the lab frame is
given by Eq.~(\ref{lortr}).

To perform the $k_0'$ integration we have to analyze the position
of the singularities in the propagators. The singularities in the
two-particle propagator $S_2(k,P')$ are  located at
\begin{eqnarray}\label{pool1}
(1) &    k_0'  = E' + E_{{\bf k}'} - i \varepsilon,
(2) &    k_0'  =  -E' + E_{{\bf k}'} - i \varepsilon,
\nonumber \\
(3) &    k_0' = E' - E_{{\bf k}'} + i \varepsilon,
(4) &    k_0' = -E' - E_{{\bf k}'} + i \varepsilon,
\nonumber \\
\end{eqnarray}
while in the initial one-particle propagator $S^{(1)} (k,P^{cm})$,
expressed in the final state variable $k'$, are given by
\begin{eqnarray}\label{pool2}
(5)         k_0' &=& (\omega^{cm} -E') + E^{cm}_{{\bf k}'-{\bf q}} - i
\varepsilon,
\nonumber \\
(6)         k_0' &=& (\omega^{cm} -E') - E^{cm}_{{\bf k}'-{\bf q}} + i
\varepsilon.
\end{eqnarray}

In the breakup region $E > M_N$ a pinching can occur between
the poles $k_0'^{(2)}$ and $k_0'^{(3)}$. This
generates the elastic cut defined by $E_{{\bf k}'} = E'$.
Moreover, the possibility exists that
the singularities $k_0'^{(2)}$ and $k_0'^{(5)}$ may cross. This
occurs when
\begin{equation}\label{crossa}
E^{cm}_{{\bf k}'- {\bf q}} - E_{{\bf k}'} = -\omega^{cm}.
\end{equation}
In Fig.~\ref{figcross} is shown the kinematic region of
$q$ and $\omega$ where this can happen.
The solid line is for small
$\omega$  given by $\omega \approx q^{lab^2}/2M_N$
which corresponds to quasielastic scattering.
Note that for real photon absorption $\omega = q$  this  double
pole doesn't occur.
The condition (\ref{crossa}) corresponds to the situation
when the hit particle before and after
the absorption of the photon is on the mass shell.
Since these poles are on the same side of the contour, their
crossing doesnot correspond to a pinching singularity.

By closing the contour in for instance the lower half plane the
integral over $k_0'$ can be performed. The integral picks up contributions
from the poles at $k_0'^{(1)}$, $k_0'^{(2)}$ and $k_0'^{(5)}$. The
singularity at $k_0'^{(1)}$ is a negative energy state contribution, the
other two are from positive energy states. In so doing we get
\begin{eqnarray}\label{contra}
I  &= & 2 \pi i \Lambda_2^-({\bf k}') S^{(2)}(k'^{(1)},P')
\tilde{\Gamma}_\mu S^{(1)}(k^{(1)},P)
\nonumber \\ & &
-2 \pi i \Lambda_1^+({\bf k}') S^{(2)}(k'^{(2)},P')
\tilde{\Gamma}_\mu S^{(1)}(k^{(2)},P)
 \nonumber \\ &&
 -2 \pi i S^{(1)}(k'^{(5)},P')S^{(2)}(k'^{(5)},P')
\tilde{\Gamma}_\mu \Lambda_1^+({\bf k})\frac{E_{{\bf k}}}{E^{cm}_{{\bf k}'-
{\bf q}}},
\end{eqnarray}
where the four vectors $k'^{(i)}$ and $k^{(i)}$ have as fourth
component $k^{(i)}_0$ and $k'^{(i)}_0$ respectively.
If we consider only positive energy intermediate states
Eq.~(\ref{int1}) reduces to
 \begin{eqnarray}\label{intpos}
 I & =& -2 \pi i \Lambda_1^+({\bf k}')\Lambda_2^+({\bf k}')
 \tilde{\Gamma}_\mu \Lambda_1^+({\bf k})
 \nonumber \\ & &
 [ \frac{1}{E+k_0^{(2)} - E_k} \frac{1}{2(E'-
 E_{{\bf k}'})} + \frac{1}{E'+ k_0'^{(5)} - E_{{\bf k}'}}
 \frac{1}{E'- k_0'^{(5)} - E_{{\bf k}'}}
 \frac{E_{{\bf k}}}{E_{{\bf k}'-{\bf q}}} ].
 \end{eqnarray}
It should be noted, that both contributions develop
separately a double pole for ${\bf k}$ satisfying
Eq.~(\ref{crossa}). However, the two terms cancel at
this point leading to only a single order pole.
Furthermore notice that only the first term has a pole
of the form $(E' - E_{k'})^{-1}$ which generates the
elastic cut.

In elastic scattering we take as integration variable $k_0$
and the the vertex $\tilde{\Gamma}_\mu $ is evaluated
in the Breit frame. In terms of the variable $k_0$ the
singularities ~(\ref{pool1},\ref{pool2}) are given by
\begin{eqnarray}\label{pool3}
(1)      k_0 &=&  E + E_{{\bf k}} - i \varepsilon,
\nonumber \\
(2)      k_0 &=& -\omega -E + E_{{\bf k}+{\bf q}} - i
\varepsilon,
\nonumber \\
(3)      k_0 &=&  E - E_{{\bf k}} + i \varepsilon,
\nonumber \\
(4)      k_0 &=& -\omega -E - E_{{\bf k}+{\bf q}} + i
\varepsilon,
\nonumber \\
(5)      k_0 &=& -E + E_{{\bf k}} - i \varepsilon,
\nonumber \\
(6)      k_0 &=& -E - E_{{\bf k}} + i \varepsilon.
\end{eqnarray}
It can readily be shown that the kinematics in this case is
such that the poles $k_0^{(2)}$ and $k_0^{(5)}$ cannot cross.
Closing the contour in the lower half plane,
the integral  becomes
\begin{eqnarray}\label{contrel}
I_{elastic}&= & 2 \pi i S^{(1)}(k'^{(1)},P')
\tilde{\Gamma}_\mu S^{(1)}(k^{(1)},P)\Lambda_2^-({\bf k})
\nonumber \\  &&
-2 \pi i S^{(1)}(k'^{(5)},P')
\tilde{\Gamma}_\mu \Lambda_1^+({\bf k})S^{(2)}(k^{(5)},P)
 \nonumber \\  &&
 -2 \pi i \frac{E_{{\bf k'}}}{E_{{\bf k}+ {\bf q}}}
\Lambda_1^+({\bf k}')
\tilde{\Gamma}_\mu S^{(1)}(k^{(2)},P)  S^{(2)}(k^{(2)},P).
\end{eqnarray}

\smallskip

\begin{figure}
\caption{The phase shifts for the BSLT
equation as a function of the lab energy for the fits given in Table I.
The solid lines are the calculated phase shifts for fit B with
the negative energy intermediate states included.
The dashed and dotted-dashed lines correspond to fits A from
Ref.\ \protect\cite{FlTj} and B respectively where only
positive energy intermediate states have been kept.
The data are from Ref.\ \protect\cite{Arndt}.
\label{nnph}}
\end{figure}

\begin{figure}
\caption{The PWIA (a and b) and the Born (c and d) contributions to the
deuteron current.
\label{pwiad}}
\end{figure}

\begin{figure}
\caption{Feynman diagrams corresponding to the IA contribution to the
deuteron current with FSI included.
\label{fsid}}
\end{figure}

\begin{figure}
\caption{The charge form factor $F_C$ and the
quadrupole form factor $F_Q$ in the relativistic impulse approximation
using the ET and the BSLT prescription for the deuteron current.
The data are from Ref.\ \protect\cite{The}.
\label{elcq}}
\end{figure}

\begin{figure}
\caption{The same as Fig.\ \protect\ref{elcq}, but for the magnetic form
factor $F_M$.
\label{elmag}}
\end{figure}

\begin{figure}
\caption{The full predictions using the ET and the BSLT
prescriptions for the deuteron current with and without MEC
contributions, but for the electric and magnetic
form factors A and B.
The data are from Refs.\ \protect\cite{elec} and \protect\cite{bib6}.
\label{elAB}}
\end{figure}

\begin{figure}
\caption{The same as Fig.\ \protect\ref{elAB}, but for the tensor
polarization $t_{20}$.
The data are from Refs.\ \protect\cite{The,Schulze,Gilman}.
\label{elt20}}
\end{figure}

\begin{figure}
\caption{The relative difference
$\frac{R^{a}_\alpha(\Gamma_{em})-R_\alpha(\Gamma_{em})}
{0.01 \times R_\alpha(\Gamma_{em})}$
of the response functions $R_\alpha$ without negative energy
state contributions and $R^{a}_\alpha$ calculated in various
approximations. When $R^{a}$ is the full result we find the solid
curve. Evaluating $R^{a}$ by neglecting both
the negative energy states and the boost transformations yields
the double dot-dash line.
Dropping furthermore also the FSI contributions in $R^{a}$
yields the long dashed curve.
The effect of the BSLT propagator choice is given by the
dashed curve. The dotted and dot-dashed curves correspond to
the results of the relative change for the various NR approximations
\protect\ref{maso} and \protect\ref{lear} to the current.
\label{relfrac}}
\end{figure}

\begin{figure}
\caption{The relative contribution of FSI to the four response
functions as function of $q$ and $\cos(\theta)$.
\label{theta_fsi}}
\end{figure}

\begin{figure}
\caption{The relative contribution of negative energy states to the
four response
functions as function of $q$ and $\cos(\theta)$.
\label{theta_neg}}
\end{figure}

\begin{figure}
\caption{The response functions for the
exclusive Nikhef kinematics. The data are from Ref.\ \protect{\cite{mvds}}.
\label{nikexpr}}
\end{figure}

\begin{figure}
\caption{Longitudinal and transverse  response functions for q =
300 MeV/c.
The data are from Ref.\ \protect\cite{Dytman}.
\label{respf1}}
\end{figure}

\begin{figure}
\caption{Longitudinal and transverse  response functions for q =
400 MeV/c.
The data are from Ref.\ \protect\cite{Dytman}.
\label{respf2}}
\end{figure}

\begin{figure}
\caption{Longitudinal and transverse  response functions for q =
500 MeV/c.
The data are from Ref.\ \protect\cite{Dytman}.
\label{respf3}}
\end{figure}

\begin{figure}
\caption{In the region I there is a crossing between two positive energy poles
of the one-particle propagator
were before and after absorption of the photon the particle is on
mass shell. In region II this crossing doesn't occur.
\label{figcross}}
\end{figure}

\end{document}